# Infographics or Graphics+Text: Which Material is Best for Robust Learning?

Kamila T. Lyra[1], Seiji Isotani[1]
Rachel C. D. Reis[1 2], Leonardo B. Marques[1],
Laís Z. Pedro[1]

[1]University of São Paulo, Brazil
[2]Federal University of Viçosa, Brazil
{kamilalyra, sisotani, rpereiramg, leobmarques, laiszagatti}@gmail.com

Patrícia A. Jaques[3],
Ig Ibert Bitencourt[4]

[3]University of Vale do Rio dos Sinos, Brazil
[4]Federal University of Alagoas, Brazil
pjaques@unisinos.br, ig.ibert@gmail.com

*Abstract*— **Infographic is a type of information visualization that uses graphic design to enhance human ability to identify patterns and trends. It is popularly used to support spread of information. Yet, there are few studies that investigate how infographics affect learning and how individual factors, such as learning styles and enjoyment of the information affect infographics perception. In this sense, this paper describes a case study performed in an online platform where 27 undergraduate students were randomly assigned to view infographics (n=14) and graphics+text (n=13) as learning materials about the same content. They also responded to questionnaires of enjoyment and learning styles. Our findings indicate that there is no correlation between learning styles and post-test scores. Furthermore, we did not find any difference regarding learning between students using graphics or infographics. Nevertheless, for learners using infographics, we found a significant and positive correlation between correct answers and the positive self-assessment of enjoyment/ pleasure. We also identified that students who used infographics keep their acquired information longer than students who only used graphics+text, indicating that infographics can better support robust learning.**

*Keywords – infographic; computer supported learning; Learning Styles; enjoyment.*

## I. INTRODUCTION

Several studies have shown that learners *remember more information* if a text is followed by key illustrations [1, 2]. Furthermore, other studies have shown that students are able to *retain information for longer* when using textual material that carries images [3]. According to the Cognitive Load Theory (CLT) [4], such phenomena happen because the use of graphics together with text reduces the cognitive load, which is the mental effort that a learner applies on learning, and thereby the learners can focus more on the content rather than trying to decode the way it is presented [4].

Thus, a positive approach to reduce cognitive load is design activities and instructional materials that match visual information, such as pictures and graphics, to textual content in order to reduce the effort spent by learners to comprehend the material [5].

In e-learning environments, there are several tools and methods to offer features that support differences in students' learning skills, learning styles and preferences [6, 7]. These possibilities bring up different ways to present and visualize educational content. Thus, many studies in the field of educational technology have investigated which is the most proper kind of visualization to enhance content assimilation [8].

More and more, information visualization is used to improve learning and knowledge retention. For example, data visualization uses graphical illustrations to communicate, in an effective way, relations between ideas and facts, and afford recognition of patterns by human cognition [9]. Other types of information visualization are used in the same direction [10]. In this context, infographic is a more recent and popular type of information visualization able to support learning. Infographics, contraction of 'information graphics', are designed to communicate a specific set of information to a certain audience by turning complex and abstract concepts into intuitive knowledge [11]. There is no restriction about the area (e.g. sciences, business) or type (e.g. numerical, data) of information that an infographic can represent. The essence/core of an infographic is in its visual appeal, blending graphical elements to represent numerical data with short and objective textual explanation. Furthermore, infographics illustrate mostly of data and textual information using icons, images, colors and elements of graphic design [12]. This appealing information visualization has the ability to guide and focus the attention of an audience, and therefore, may be used as learning material with the potential to increase knowledge acquisition.

Although the popularity of infographics is unquestionable, we observe a lack of researches that verify its effectiveness as learning material. In this context, we investigate how learning and information retention is affected by infographics when using them as educational material. Furthermore, since infographics use graphic design elements to produce very appealing visualizations, we also explore whether exists a correlation of Learning Styles [13] and students' satisfaction and enjoyment [14, 15] with learning using infographics. In order to collect information about the influence of this kind of visualization, we conducted a randomized controlled study with 27 engineering undergraduate students to understand whether instructional materials presented as infographics can be more effective for learning than conventional instructional materials presented using a combination of graphics and texts (graphics+text).

This paper presents other three following sections. In Section 2 we present the related works. In Section 3 we detail the conduction of the randomized controlled study. In Section 4 we present the findings from the collected data and discuss the analysis. In Section 5 we have the conclusions.

## II. RELATED WORK

In order to obtain an overview of scientific studies related to infographics and education, we conducted a systematic mapping of literature [1]. Even though the popularity of infographics, our review found few studies that verify its effectiveness as learning material concerning short-term memory and retention.

Some of the studies that apply infographics as learning material report teachers' experience on using infographics design as a template [17, 18]. The students described the class content by using an infographic to expose their understandings. Although the authors compliment this method, these studies did not perform any test to verify students learning, also they did not compare learning by infographics with other types of educational material.

Zhang-Kennedy et al. [19] and McTigue [20] tested students to compare learning of those who used infographics to learning of students using pure text. Both studies identified greater results for learners who used infographics. Students who used pure text had little gain regarding learning, information retention and usefulness of the information. Despite the evaluation, these studies did not concern about individual factors that may interfere in how students perceive the learning material.

Diakopoulos et al. [21] explored the use of infographics on learning adding levels of game elements. They evaluate and discuss infographics concerning insights and learning. Moreover, this work cares about how much students enjoy the activity. In a questionnaire, the researchers asked the participants to evaluate how enjoyable and fun was their experience on a Likert scale. However, the students' rating was concerning the various degrees of game elements instead of material presentation.

Ultimately, the above references characterize the research field about 'infographics in education'. Still, few works provide evidences on how the appealing design of an infographic can affect learning process, and whether individual factors such Learning Styles and enjoyment of the information interpose on those evidences.

## III. METHOD

### A. Participants

A class of 27 Environmental Engineering undergraduate students was recruited to perform the study. Since infographics can merge text, images, graphics and design to represent information, we selected undergraduate students to grant the ability to interpret infographics holistically. The ability to read multiple sources of information represents a higher order comprehension skill that is a hallmark of adult readers [22]. Young learners are more likely to observe an isolated component rather than considering the visualization holistically [23].

### B. Learning Material

We selected 15 infographics from varied topics since we did not want the influence of student's specific knowledge. So we opted for common knowledge themed infographics such as energy saving, environment and recycling. Fig. 1 shows an example of a used infographic.

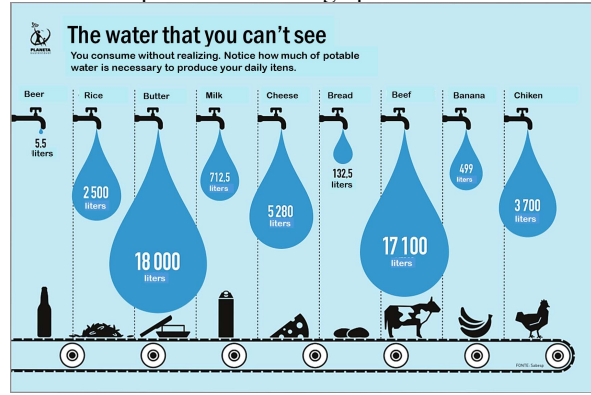

Figure 1. Infographic example.

To verify whether the appealing design of an infographic impacts learning, we converted the infographics into graphics+text (text blended to simple graphics). All illustration and design elements were removed and all visual information was transcribed. However, we concerned to keep graphical representation to illustrate data. This process was executed to ensure that the text plus simple graphic configuration maintained exactly the same information and data from the infographics. The resulting representation from the infographic in Fig. 1 is shown in Fig. 2. Finally, for our study, we used 15 infographics and their graphics+text versions (also 15) as learning material.

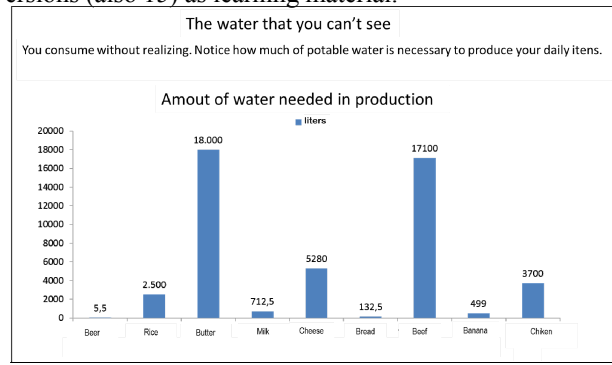

Figure 2. Conversion to graphic+text.

### C. Online platform

The online environment in which students performed the experiment was developed using Java programming language. We used the Model-View-Controller (MVC)

---
[1] A systematic mapping is a sort of literature review that uses a well-defined methodology to identify, analyze and interpret the studies [16].

VRaptor framework and the Hibernate framework for the connection with MySQL database. We used Tomcat server and the software architecture followed MVC standard.

*D. Study design*

We performed a study to verify if the design employed by infographics can affect learning process. This study aims to compare learning and information retention from learners using infographic to graphic+text. Students passed by three stages, one week apart each: in Stage A they answered a pretest questionnaire (45 questions) and a learning style questionnaire (20 questions) [24]; in Stage B students viewed the learning material and answered a post-test questionnaire (45 questions); at last, Stage C presented the delayed post-test questionnaire (45 questions). All stages were performed in a computer lab session using the developed online platform.

*E. Stage A*

Before viewing the learning material, the 27 participants answered the pretest questionnaire to provide us with their initial knowledge about the content. This initial value was used as base to compare the students' knowledge before and after viewing the learning material. The pretest is composed of 45 multiple choice questions about data that are explicitly stated on the selected infographics. Therefore, each infographic generated 3 questions, affording 4 options of answers: the correct answer, two wrong answers and "I don't know" option. This last one was placed on the test attempting to decrease the occurrence of guessing. Still in Stage A we collected answers to determine student's Learning Styles. The applied questionnaire (New-ILS) is a version of the Index of Learning Styles by Felder and Soloman [25][2].

*F. Stage B*

In this Stage students interacted with the learning material; they were assigned to one of the two versions of the system. Both systems implement the same features; they only differ on the learning material type. One system presented 15 infographics, while the other displayed the 15 corresponding conversions to graphic+text. Thereby 14 students used infographics and 13 used text. Fig. 3 illustrates the flow of tasks in Stage B.

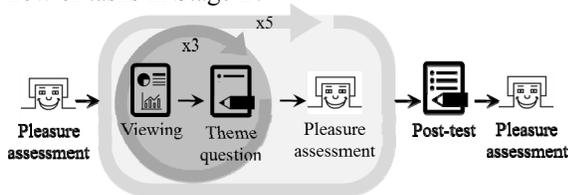

Figure 3.  Tasks in Stage B.

Students were asked to state their satisfaction with the activity using the Self-Assessment Manikin (SAM) pleasure scale [26] in three moments: before starting using the system, after each cycle of visualization (they were three in total), and after the post-test questionnaire. Fig. 4 presents the SAM pleasure dimension. It depicts five graphic characters, from smiling to a frowning, displayed along a nine point scale, in which 1 denotes very bad and 9 very well. Thus, both groups settled their satisfaction with the activity seven times spread over the Stage B.

---

[2] This version is translated to Portuguese (Brazil), reduced from 44 original questions to 20 questions and was validated [24].

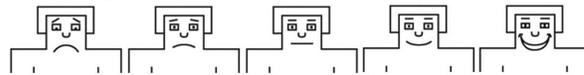

Figure 4.  SAM pleasure dimension.

After every visualization, participants answered one question about what was the material themed. The theme question was applied as an alternative mean to encourage students to read the complete visualization. The viewing cycle was complete when students passed through 15 visualizations, for either infographic or graphic+text versions. Right after the viewing cycle, students answered a post-test questionnaire composed with the same 45 questions from the Pretest in a rewritten vocabulary and disposed in different order. Participants settled their satisfaction one last time as a final task for this stage.

*G. Stage C*

The last stage was a delayed post-test questionnaire applied one week after Stage B. The purpose of this questionnaire was to assess information retention from post-test to this stage and evaluate whether there were improvements in knowledge. The last Stage questionnaire was composed by the same 45 questions from pretest and post-test questionnaire; again disposed and written differently from the previous stages.

IV. ANALYSIS AND DISCUSSION

Observing the data of all students collected in all three Stages (A, B, and C) presented in Section 3, we can confirm the intervention as a good instructional setting. A paired samples t-test compared the number of correct answers between the pretest and the post-test and found a statistically significant difference ($t(52) = -10.9$; $p < .001$) between the mean of correct answers in the pretest ($\mu = 20.85$, $\sigma = 3.5$) and the mean in the posttest ($\mu = 33.96$, $\sigma = 5.11$). This significant difference shows the effectiveness of the experiment design, since students scored more after interacting with the learning material, regardless of what was the material type. Despite the slight average decrease from the post-test to the delayed post-test ($\mu = 31.44$, $\sigma = 6.55$), the difference was not statistically significant ($t(52) = 1.5$; $p = .12$). Fig. 5 exhibits the average number of correct answers in each Stage. Thus, we can say that learners maintained the acquired knowledge, even though one week passed by since intervention.

We observed similar findings when we compared students by the type of the learning object used in the intervention. For Stage B, 13 students viewed graphic+text while 14 students used infographics. As shown in Table I, a significantly higher score mean was found in the posttest both for the infographics condition (t(26) = -6.8, p < .001) and for the graphic+text condition (t(24) = -9.02, p < .001). However, the difference of the mean of the post-test and the delayed post-test was not statistically significant for both groups.

Although statistics showed little variation between learning by graphic+text or infographic, it is possible to compare the increase and decrease value from one stage to another and between groups (graphic+text and infographic). Table I contains the average number of correct answers separated by learning material in each Stage. At all Stages, students who used infographics during intervention scored higher in average than students that viewed graphic+text, though this difference is not statistically significant. Despite this, subjects using graphic+text showed a higher gain from pretest to post-test and, at the same time, a higher decrease from post-test to delayed post-test, as shown in Fig. 5. Thus, even if students acquired more knowledge using graphic+text, this type of visualization was not as effective as infographics regarding knowledge retention.

TABLE I.     AVERAGE NUMBER OF CORRECT ANSWERS BY STAGE AND LEARNING MATERIAL

| Stage | Learning Material | |
|---|---|---|
| | *Infographic* | *Graphic+text* |
| Pretest | 22.07 | 19.54 |
| Post-test | 34 | 33.92 |
| Delayed post-test | 32.07 | 30.77 |

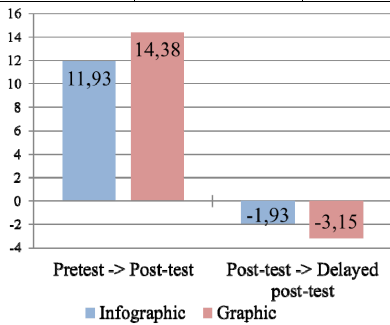

Figure 5.   Difference between Stages

The purpose of including pleasure self-report data obtained during the 'viewing and post-test' stage (Stage B) was to explore whether the learning material can affect student's sense of enjoyment and also if this feeling had an effect on the post-test performance. Data showed no statistically significant difference between the average pleasure given by students using graphic+text ($\mu = 5.24$) and the average by students using infographics ($\mu = 5.56$) though the last one was slightly higher[3]. Thereby, it is not possible to say which material neither how (positively or negatively) it affected the declaration of enjoyment.

---

[3] Five is the midpoint in the Self-Assessment Manikin scale [27].

However, we also questioned whether this pleasure declaration was connected to the number of correct answers in the post-test questionnaire. For subjects using infographics as learning material, we found a significant and positive correlation ($p < .05$) between the number of correct answers and the positive self-assessment of pleasure. In other words, the higher a student scores in post-test, more likely s/he is to indicate a positive satisfaction. In contrast, we did not found the same significant correlation for students who used graphic+text as learning material. Thus, it is not possible to confirm the correlation between scores in post-test and self-assessment of pleasure to this group.

We also analyzed whether the positive self-assessment of pleasure was related to a combination of Learning Styles and learning material. For the group of students using *infographics*, the New-ILS questionnaire identified 10 visual learners and 4 verbal learners. Table II shows that verbal students on average set a higher value in the pleasure scale than visual learners. The questionnaire classified 8 students as visual and 5 as verbal learners in the group that used graphic+text. In this group, visual students had a higher mean in the pleasure scale than verbal students. Despite the difference between the averages, it is not possible to say that visual learners enjoyed more using graphic+text than infographics, since t-test did not reveal a significant difference. As students were randomly assigned to one of the materials, the number of visual and verbal students differs in each group. Thus a larger and numerically similar sample can provide us more precise evidence about how learning styles affect the pleasure self-assessment according to the type of the learning material.

TABLE II.     AVERAGE PLEASURE BY LEARNING STYLE

| Learning Style | Learning Material | |
|---|---|---|
| | *Infographic* | *Graphic+Text* |
| Visual | 5.2143 | 5.7679 |
| Verbal | 6.4286 | 4.4000 |

Moreover, in our data analysis we did not find evidence that learning styles have an effect on the number of correct answers. No learning style dimension affected the number of correct answers in post-test for the group of students who used graphic+text, neither for the infographics users. This finding follows what some researchers have discussed about the lack of evidences that Learning Styles benefits learning [27, 28]. Data presented here reveal that self-assessment of pleasure was correlated to learning in a significant measure, while learning styles did not affect it. Thus, in the context designed for this study, factors such as enjoyment might be more impactful than learning styles for choosing a learning material to improve students' learning.

## V. CONCLUSION

Computers are extensively used to support learning process. It can offer features capable to overlap differences in learning skills by adapting the format of the material to the learners needs. However, education researchers still investigate which is the most proper kind of visualization to use in computer supported learning in order to enhance understanding. In this context, this paper presents a case study developed in an online platform to verify whether the design employed by infographics can affect learning, knowledge retention and self-assessment of enjoyment.

Regarding learning, our study did not show statistical difference between students using graphic+text and students using infographics. Although the average number of correct answers of students that used graphic+text increased more from pretest to post-test, the average number of correct answers from post-test to delayed post-test also decreased more than the students that used infographics. This can be an evidence that infographics are more effective concerning knowledge retention.

Self-assessment of pleasure seems not to be affected when comparing students by learning material (infographics and graphic+text). Yet, for students using infographics the value given for enjoyment was correlated to the score reached in post-test. The higher the score, more positive were the pleasure assessment.

We observed Learning Styles in order to catch evidences to prove the preferences of visual or verbal learners for certain type of learning material. However, our analysis did not show statistical evidences that visual or verbal learners settled more positive values for pleasure to any of the learning materials. Likewise, learning styles did not show influence concerning the number of correct answers in post-test for any group. Thus, in the design applied for this study, we found that factors such as enjoyment have a greater influence on learning than learning styles. As future research, we plan to replicate this experiment using a larger sample of students.

ACKNOWLEDGEMENTS

The authors thank CNPq for its financial support.